\documentclass[twocolumn,showpacs,floatfix,bibnotes]{revtex4}

\usepackage{graphicx}

\newcommand\pictc[5]{\begin{figure}
                       \centerline{\vspace{0mm}
 \includegraphics[width=#1\columnwidth,height=0.7\textheight,keepaspectratio]{#3}}
                       \protect\caption{\protect\label{fig:#4} #5}\vspace{0mm}
                    \end{figure}            }
\newcommand\pict[4][1]{\pictc{#1}{!tb}{#2}{#3}{#4}}
\newcommand\rpict[1]{\ref{fig:#1}}

\begin{document}
\begin{sloppy}

\title{Polychromatic interface solitons in nonlinear photonic lattices}

\author{Kristian Motzek}
\author{Andrey A. Sukhorukov}
\author{Yuri S. Kivshar}

\affiliation{Nonlinear Physics Centre and Centre for Ultra-high
bandwidth Devices for Optical Systems (CUDOS), Research School of
Physical Sciences and Engineering, Australian National University,
Canberra ACT 0200, Australia}

\begin{abstract}
We demonstrate that interfaces between two nonlinear periodic
photonic lattices offer unique possibilities for controlling
nonlinear interaction between different spectral components of
polychromatic light, and a change in the light spectrum can have a
dramatic effect on the propagation along the interface. We predict
the existence of polychromatic surface solitons which differ
fundamentally from their counterparts in infinite lattices.
\end{abstract}

\maketitle

Nonlinear optics has focused on the study of self-action of
monochromatic light for many decades. This was due to the fact that
high-power light necessary to observe strong nonlinear effects could
only be obtained from laser sources, which usually show only a few
rather narrow spectral lines. However, since nonlinear photonic
fibers have been used to successfully generate supercontinuum
radiation, polychromatic light is attracting more and more attention
in the nonlinear optics community. After an early paper on nonlinear
focusing of white light and incoherent spatial
solitons~\cite{MitchellNature97}, several papers studied the topic
of spatially localized modes for polychromatic light in nonlinear
media~\cite{BuljanOL03,Motzek:2005-WD39:ProcNLGW}. As a general
result, we mention that it has always been observed that the blue
components of a polychromatic light beam were much better localized
than the red components, because the diffraction is weaker for light
of shorter wavelength.

In this Letter, we study the light localization near the interface
between two nonlinear semi-infinite periodic photonic lattices and
the generation of polychromatic interface solitons. In particular,
we show that nonlinear interfaces can be tailored to obtain the
opposite result: self-focusing of the red parts of the spectrum with
the blue parts which only weakly or even not at all localized.
Furthermore, we show that at interfaces the nonlinear interaction
between spectral components can have dramatic effects on the
propagation of light offering a new approach to control the spectrum
of polychromatic light.

In many recent studies, the properties of light
propagating through periodic photonic lattices have been explored.
The bandgap structure of the lattice spectrum plays the
decisive role for many intriguing effects observed. Also, the
interfaces between a photonic lattice and homogeneous media have
been known to be of particular interest for a long time, because
they support linear modes localized at the surface, the so-called
optical Tamm states~\cite{Yeh_APL_78}. Recently, it was predicted
theoretically and demonstrated experimentally that nonlinear
self-trapping of light near the edge of a waveguide array which can
lead to the formation of discrete surface
solitons~\cite{OL_surface,Suntsov:2006-063901:PRL}, surface gap
solitons~\cite{KartashovPRL06}, and multi-gap surface
solitons~\cite{surface3}. This concept can also be extended to
interfaces between two different photonic lattices. Obviously, for
any localization to be possible, there has to be an overlap of the
bandgaps of the lattices.

\pict[0.95]{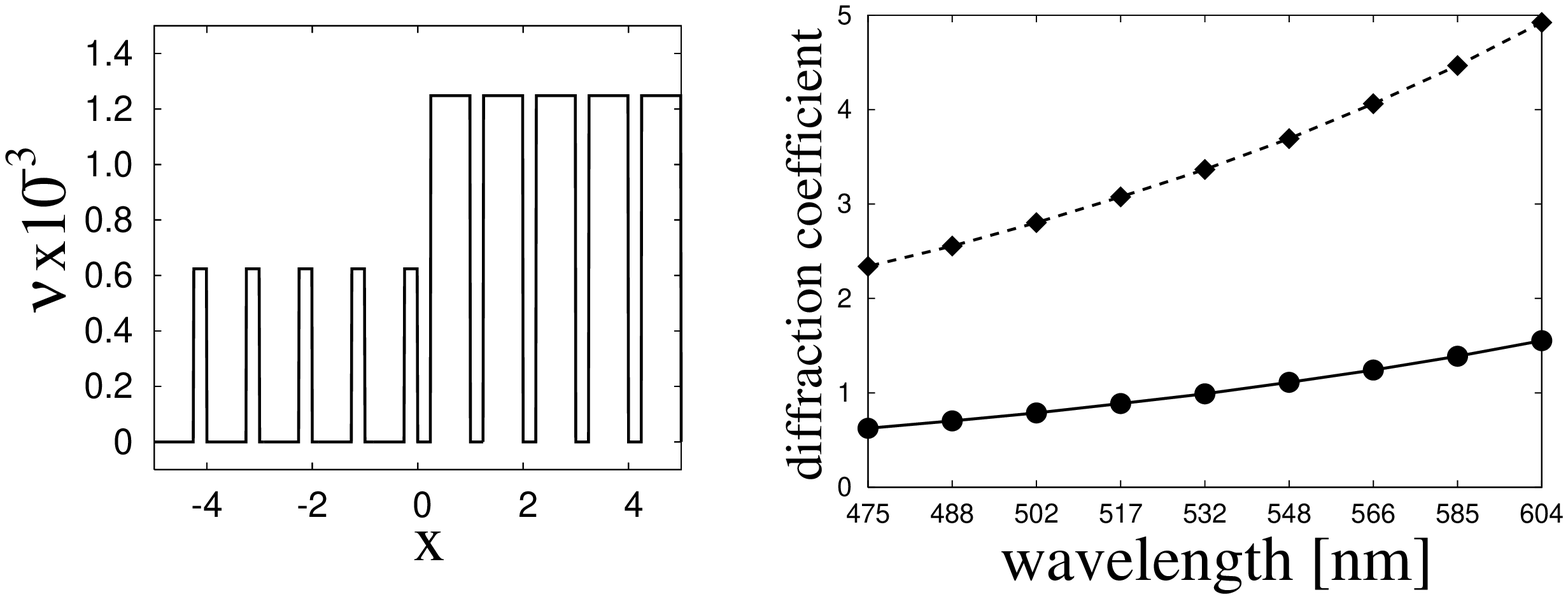}{Lattice}{ (a) Schematic of an interface separating
two photonic lattices. (b) Diffraction coefficients for the bottom
of the first band of the narrow- (solid line) and the second band of
the wide-waveguide lattice (dashed line).}

We consider the
interface between two periodic photonic lattices shown in
Fig.~\rpict{Lattice}(a). Both lattices to have the same
period but having different unit cells. In the region $x<0$ the
lattice consists of narrow waveguides, whereas for $x\ge 0$ we
choose the waveguides to be wider and deeper.
Figure~\rpict{Lattice}(b) shows the diffraction coefficient (i.e.
the curvature of the spectral bands) at the bottom of the first band
of the narrow-waveguide and at the bottom of the second band
of the wide-waveguide lattice for different wavelengths. We observe
that the blue spectral components could be localized much easier
than the red ones if they were propagating within one of the
lattices.

We study the propagation of polychromatic light beams described by
the system of coupled equations
\begin{eqnarray}
\label{PropagationEq}
i\frac{\partial A_n}{\partial z} + \frac{\lambda_n z_0}{4\pi n_0
  x_0^2} \frac{\partial ^2 A_n}{\partial x^2} + \frac{2\pi
  z_0}{\lambda_n} \left[\nu(x) - \gamma I\right] A_n = 0 \, ,
\end{eqnarray}
where $A_n$ are the envelopes of different wavelength components of
vacuum wavelength $\lambda_n$, $\nu(x)$ stands for the periodic
modulation of the refractive index in the transverse spatial
dimension, $I = \sum_n |A_n|^2$ is the total light intensity, and
$\gamma$ measures the nonlinearity strength. In
Eqs.~(\ref{PropagationEq}), the transverse ($x$) and longitudinal
($z$) coordinates are normalized to $x_0 = 10\mu$m and $z_0 = 1$mm,
respectively, and the nonlinearity is defined by $\gamma = 10^{-3}$.

\pict{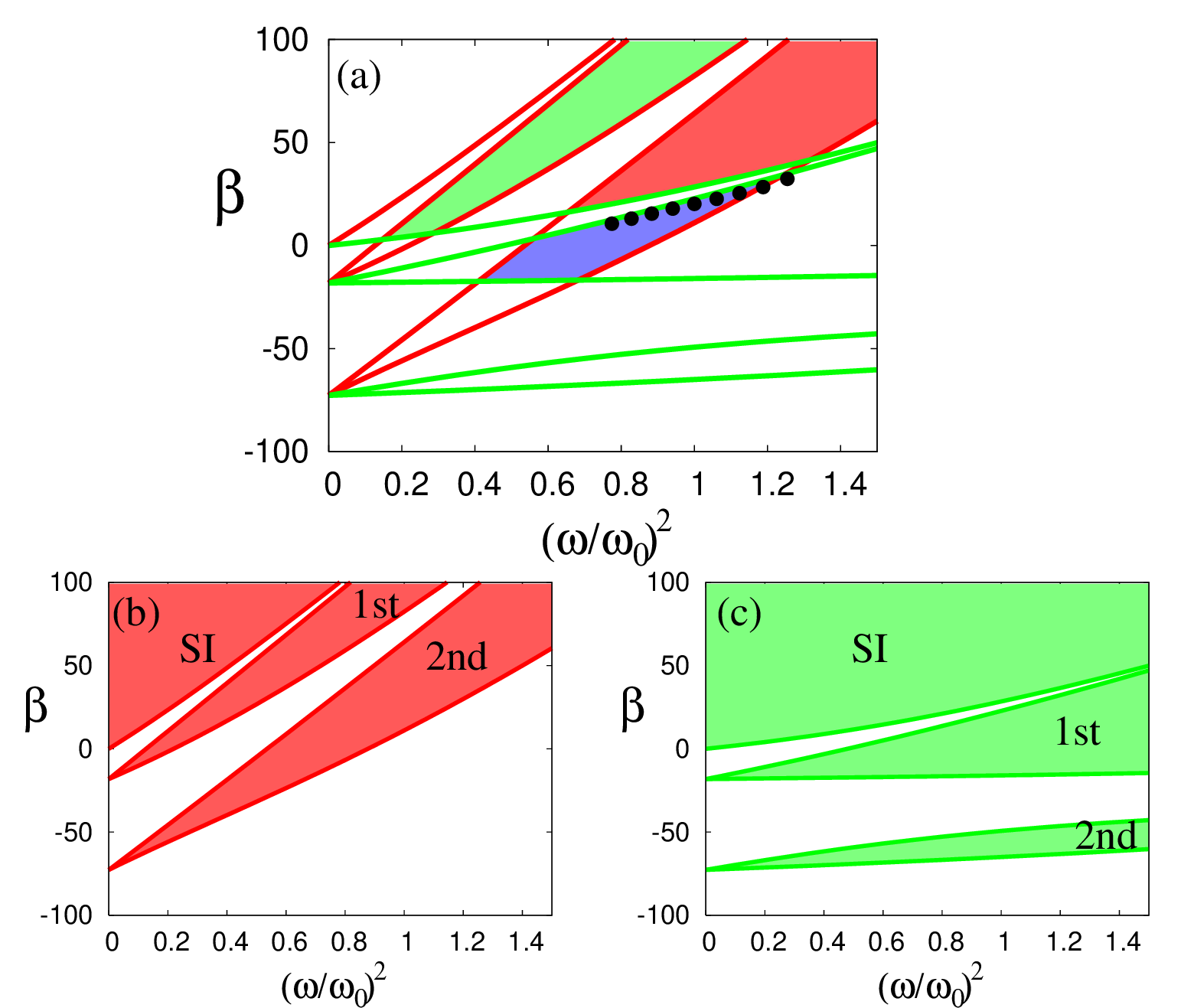}{BandGaps}{ (Color online) (a-c) Bandgap spectra of the separate and
combined photonic structures. (b,c) Bandgap spectra (shaded) of the
wide- and narrow-waveguide lattices vs. frequency, respectively. and
(c) their overlaps. In (c) the darkest shaded region is the overlap
between the second gap of the wide-waveguide lattice and the first
gap of the narrow-waveguide lattice.}

The bandgap structure of both lattices as a function of the light
frequency (scaled to $\omega_0=2\pi c/532$nm) is shown in
Figs.~\rpict{BandGaps}(b,c) while (c) shows their overlaps. We chose
the lattices in such a way that the overlap between the first gap in
the region $x<0$ and the second gap in the region $x \ge 0$ vanishes
above a certain cut-off frequency $\omega_c\approx
2\pi c/475\mbox{nm}\approx \sqrt{1.25} \omega_0$.

First, we analyze the existence of polychromatic localized modes at
the interface, i.e. {\em polychromatic surface solitons}. We choose
the frequency range close to the cut-off frequency $\omega_c$, and
find numerically different types of localized modes.
Figure~\rpict{Soliton} shows an example where the polychromatic
interface soliton is composed of five components with different
wavelengths $\lambda=506, 519, 532, 546$ and $560$nm (equidistantly spaced in frequency space). All five
components carry the same power. In the soliton, the blue components
have a larger spatial extent that the red ones. This is in a sharp
contrast to other types of polychromatic solitons in infinite
systems.

\pict[0.85]{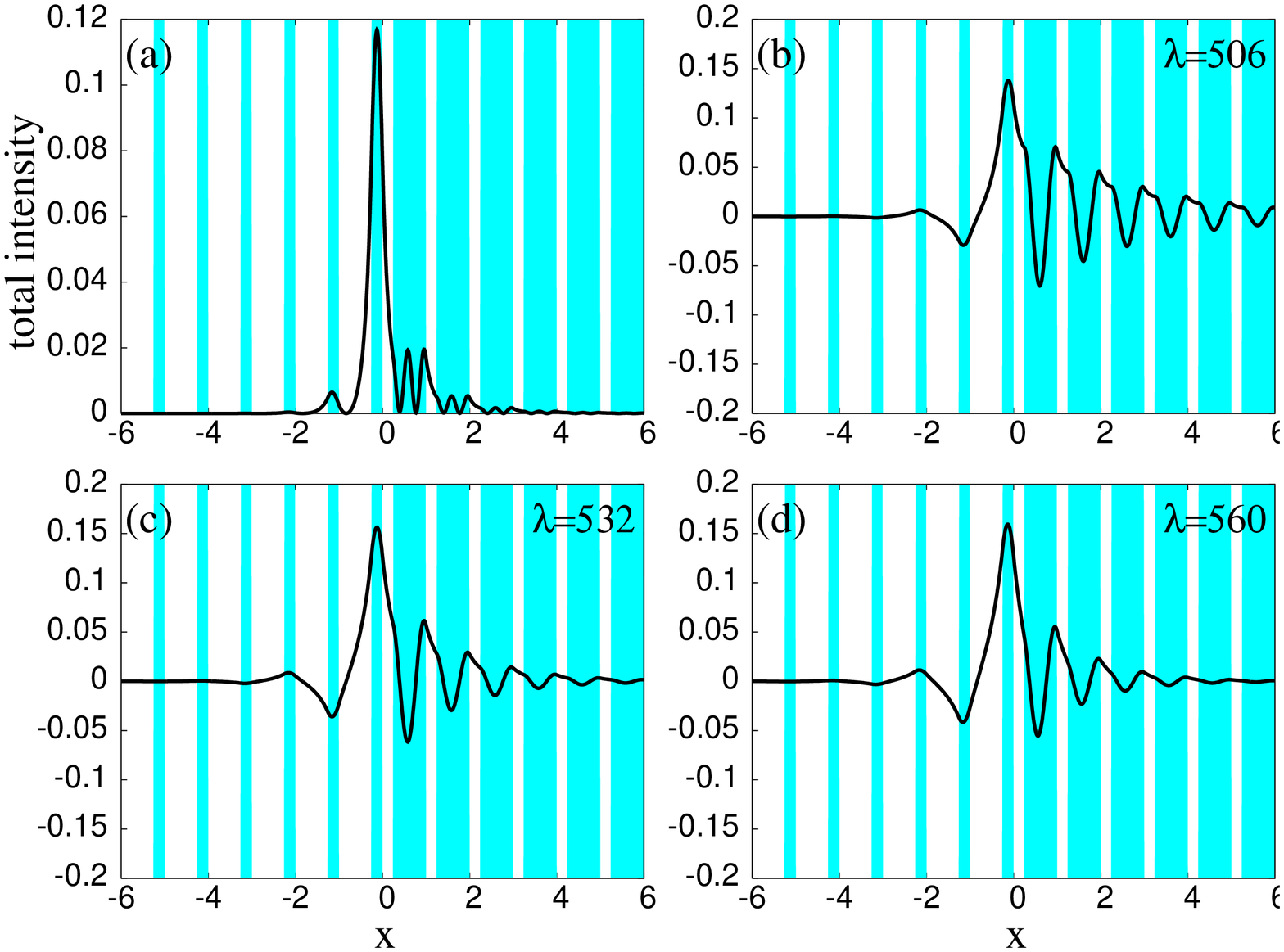}{Soliton}{An example of a polychromatic surface
soliton consisting of five components of different wavelengths.
Shown are the total intensity and three components.}

Single components shown in Fig.~\rpict{Soliton} are indeed located
within the first spectral gap of the narrow-waveguide lattice and
the second gap of the wide-waveguide lattice. We study numerically
the propagation of such multi-component soliton in the presence of
an initial perturbation and could not observe any signs of
instabilities. This is in agreement with results obtained for
monochromatic surface solitons~\cite{KartashovPRL06}.

However, polychromatic surface solitons are of practical relevance
only if they can be generated under experimentally realistic
conditions. Therefore, we simulate numerically a situation where
polychromatic light with a Gaussian intensity profile is injected
into the narrow waveguide closest to the interface. The
polychromatic light is modeled by nine components with different
wavelengths $\lambda=$475, 488, 502, 517, 532, 548, 566, 585, and
604nm. Our results clearly show that it is indeed possible to
generate a polychromatic surface soliton in this way.
Figure~\rpict{Generation}(a) shows the evolution of the total
intensity with propagation and Fig.~\rpict{Generation}(b) shows the
spectrum of the light beam at the input and output (considering only
the waveguides closest to the interface and the space between them).
We chose the light to have a flat spectrum at the input.
Fig.~\rpict{Generation}(b) shows that the propagation along the
interface considerably alters the beam spectrum. Most of the
intensity of the red part of the spectrum is trapped into the
interface soliton, while most of the blue part diffracts away from
the interface. This is due to the fact that we chose the
polychromatic light beam to lie in a frequency range close to the
cut-off frequency $\omega_c$, above which the overlap between the
first gap of the narrow- and the second gap of the wide-waveguide
lattice disappears.

\pict[0.9]{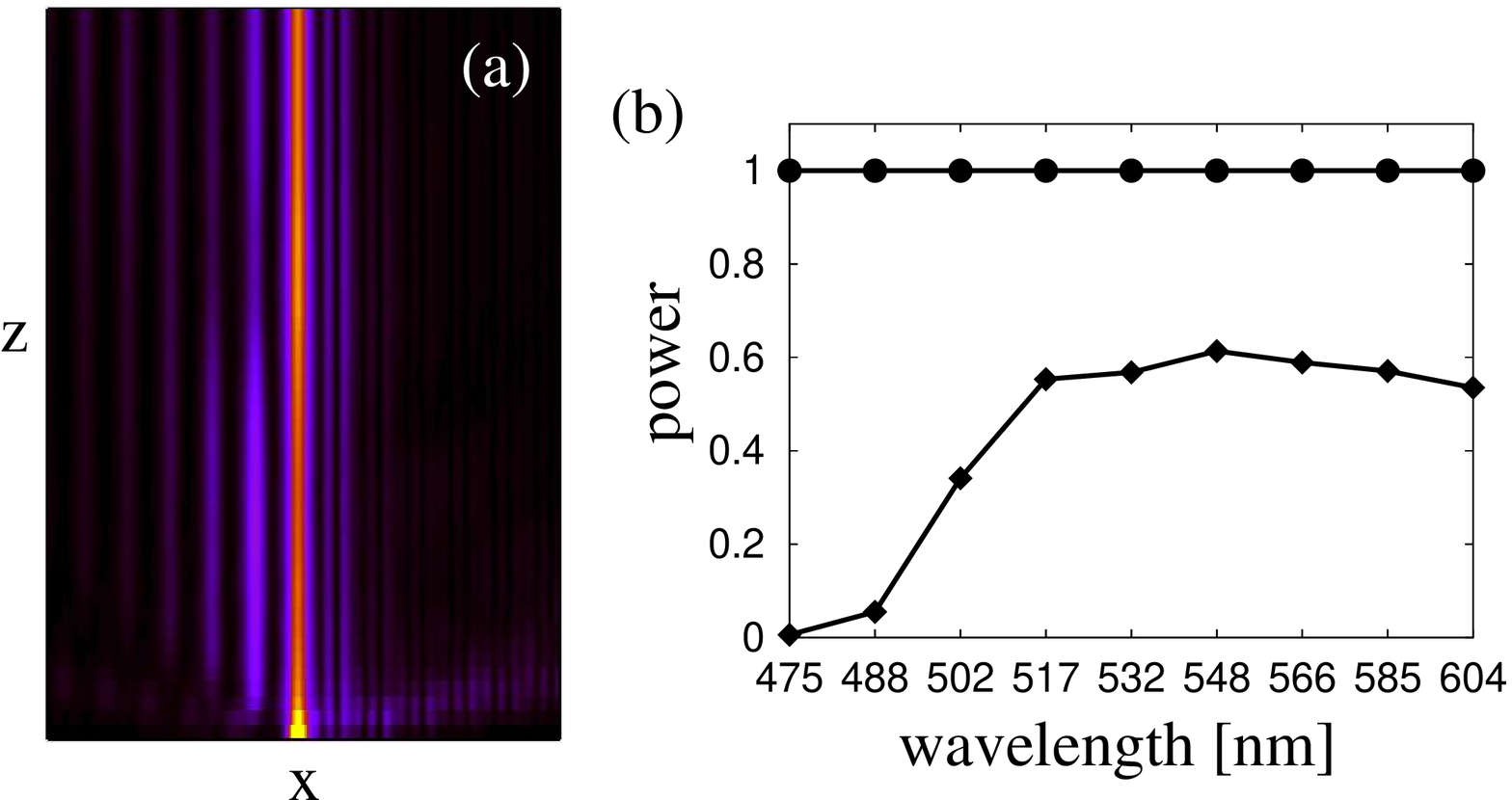}{Generation}{(Color online) Generation of a polychromatic interface
soliton. (a) Evolution of the total beam intensity over 10~cm
propagation (bottom to top), (b) input and output spectra.}

However, we have seen in Figure~2(c) that there is not only an
overlap between the first gap of the narrow- and the second gap of
the wide-waveguide lattice. The semi-infinite gap of the narrow
overlaps with the first and second gap of the wide-waveguide lattice
in the frequency region under consideration. Surface solitons should
exist in these band-gap overlaps as well. To understand, why the
soliton is forming not in these overlaps, but in the overlap between
the spectral gaps, one has to keep in mind that in an infinitely
extended medium (i.e. in the absence of any interface) solitons can
form in the semi-infinite gap only in the case of a self-focusing
nonlinearity. Here, however, we have a self-\emph{de}focusing
nonlinearity. Let us now consider a surface soliton that resides in
a spectral gap of one and the semi-infinite gap of the other
lattice. In the case of defocusing nonlinearity, increasing the
soliton intensity will make localization in the semi-infinite gap
more difficult, because diffraction is enhanced by nonlinearity.
Increasing the intensity further, the
nonlinearity-enhanced diffraction in the semi-infinite gap is too
strong for any surface state to exist. Therefore 
the light beam excites preferentially a soliton in the overlap
between the spectral gaps of the two lattices.

\pict[0.9]{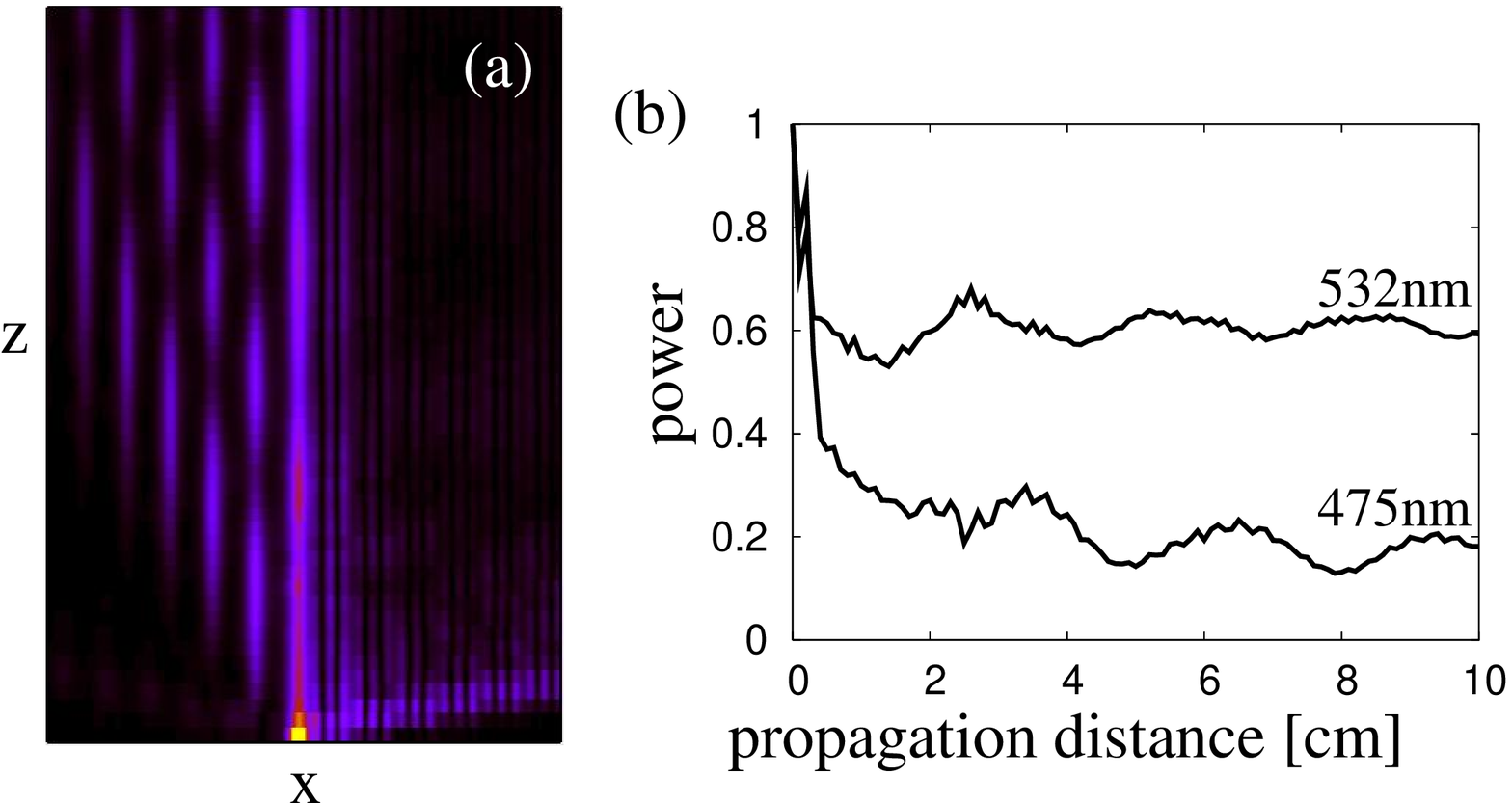}{Monochromatic}{(Color online) Propagation of a monochromatic light
beam along the interface. Except for the spectrum all parameters are
identical to those in Fig.~\rpict{Generation}. (a) Evolution of the
total intensity over 10~cm propagation; (b) Amount of the power
localized at the interface for 475~nm and 532~nm.}

For the blue components this overlap is too small and they
diffract. The situation changes though, when we look at the
propagation of a monochromatic light beam at the blue end of the
spectrum. In the absence of the red part that did excite a soliton
in our polychromatic simulation, the blue light is free to go into
the overlap between the semi-infinite gap of the narrow and the
second gap of the wide waveguide lattice. (Localized solutions
do exist in this overlap despite the nonlinearity enhanced diffraction
in the semi-infinite gap.)
This is seen in
Fig.~\rpict{Monochromatic}, which shows the propagation of a
monochromatic light beam of wavelengths $\lambda=475\mbox{nm}$.
Except for the spectrum, all parameters are the same as in
Fig.~\rpict{Generation}. In Fig.~\rpict{Monochromatic}(a) we observe
that a noticeable fraction of the power remains at the interface.
This is quantified in Fig.~\rpict{Monochromatic}(b), and also
compared to the case of $\lambda=532\mbox{nm}$, showing the
evolution of the power localized at the interface as a function of
propagation distance. After 10cm of propagation roughly 20\% of the
initial power is still remains at the interface. This is a striking
result, when comparing it to the results of the polychromatic
soliton generation, where basically all the power of the 475nm
component has diffracted away from the interface after 10cm of
propagation, as can be seen in Fig.~\rpict{Generation}(b).
Thus we have a situation in which the presence of the other (localized)
components prevents the localization of the 475nm beam. 
We note that the opposite
effect (enhanced localization due to the interaction with other
components) can also be observed in our system when moving to the
other end of the overlap between the band-gaps.

The different behavior of the $\lambda=475$nm
component in the poly- and the monochromatic case highlights the
intriguing nonlinear interaction between the spectral components of
polychromatic light propagating along a nonlinear interface. The
complex nature of the interaction leaves much space for engineering
interfaces that can transform the spectrum of polychromatic light in
a specific way.

In conclusion, we have predicted the existence of polychromatic
interface solitons localized at the interface separating two
different semi-infinite periodic photonic lattice, and demonstrated
that such multi-component gap solitons can differ considerably from
their counterparts in infinite photonic lattice. In particular, our
study of the localization of polychromatic light near the interface
has demonstrated that the red parts of the spectrum can be localized
better than the blue parts, so that the interface can be employed
for controlling nonlinear interaction between spectral components of
polychromatic light.

\end{sloppy}
\end{document}